\documentclass{pasj00}

\def\PublisherReidel{Dordrecht: D. Reidel Publishing Company}

\begin{document}
\SetRunningHead{M. Uemura et al.}{Early Superhump Mapping}
\Received{2011/12/27}
\Accepted{2012/3/6}
\Published{}

\title{Reconstruction of the Structure of Accretion Disks in Dwarf Novae from the
  Multi-Band Light Curves of Early Superhumps}

\author{Makoto \textsc{Uemura}} 
\affil{Hiroshima Astrophysical Science Center, Hiroshima University,
  Kagamiyama 1-3-1, Higashi-Hiroshima, Hiroshima 739-8526}
\email{uemuram@hiroshima-u.ac.jp}

\author{Taichi \textsc{Kato}, Tomohito \textsc{Ohshima}}
\affil{Department of Astronomy, Kyoto University,
  Kitashirakawa-Oiwake-cho, Sakyo-ku, Kyoto 606-8502} 
\and
\author{Hiroyuki \textsc{Maehara}}
\affil{Kwasan and Hida Observatories, Kyoto University, Yamashina-ku,
  Kyoto 607-8471} 

%


\maketitle

\begin{abstract}
We propose a new method to reconstruct the structure of accretion
disks in dwarf novae using multi-band light curves of early
superhumps. Our model assumes that early superhumps are caused by
the rotation effect of non-axisymmetrically flaring disks. We have
developed a Bayesian model for this reconstruction, in which a
smoother disk-structure tends to have a higher prior probability. We
analyzed simultaneous optical and near-infrared photometric data of
early superhumps of the dwarf nova, V455~And using this technique. The
reconstructed disk has two flaring parts in the outermost region of
the disk. These parts are responsible for the primary and secondary
maxima of the light curves. The height-to-radius ratio is
$h/r=0.20$---$0.25$ in the outermost region. In addition to the 
outermost flaring structures, flaring arm-like patterns can be seen in
an inner region of the reconstructed disk. The overall profile of the
reconstructed disk is reminiscent of the disk structure that is
deformed by the tidal effect. However, an inner arm-like pattern,
which is responsible for the secondary minimum in the light curve,
cannot be reproduced only by the tidal effect. It implies the presence
of another mechanism that deforms the disk structure. Alternatively,
the temperature distribution of the disk could be non-axisymmetric. We
demonstrate that the disk structure with weaker arm-like patterns is
optimal in the model including the irradiation effect. However, the
strongly irradiated disk gives quite blue colors, which may conflict
with 
the observation. Our results suggest that the amplitude of early
superhumps depends mainly on the height of the outermost flaring
regions of the disk. We predict that early superhumps can be detected
with an amplitude of $>0.02$~mag in about 90~\% of WZ~Sge stars. 
\end{abstract}

\section{Introduction}

Early superhumps are periodic variations occasionally observed in
WZ~Sge-type dwarf novae in their superoutbursts
(\cite{kat96alcom}). Their period is extremely close to the orbital
period of the binary system. Their light curve has a double-peaked
profile with different amplitudes. These characteristics of early
superhumps suggest that they are caused by a different mechanism from
that of ordinary superhumps, which are commonly observed in all
SU~UMa-type dwarf novae; the period of ordinary superhumps is a few
percent longer than the orbital period, and the light curve generally
has a saw-tooth profile with a prominent single peak
(\cite{vog80suumastars}; \cite{war85suuma}; for a recent review, see
\cite{kat09pdot}). The amplitude of early superhumps is $\lesssim
0.5$~mag, and is largest at the superoutburst maximum. Then, early
superhumps decreases in amplitude with time, and finally, ordinary
superhumps supersede early superhumps $\sim 10$~d after the
superoutburst maximum (\cite{ish02wzsgeletter};
\cite{kat02wzsgeESH}). Thus, early superhumps are observed only in the
early stage of superoutbursts of WZ~Sge stars.

The mechanism of early superhumps is poorly understood. In contrast,
the mechanism of ordinary superhumps have been understood in terms of
the tidal dissipation in eccentric, precessing accretion-disks. A disk
in a binary system is theoretically expected to have an eccentric form
when it expands beyond the 3:1 resonance radius (\cite{whi88tidal};
\cite{lub91SHa}). In the case of WZ~Sge stars, the disk possibly
expands far beyond the 3:1 resonance radius. This is due to their
extreme mass-ratios ($M_2/M_1\lesssim 0.1$, where $M_2$ and $M_1$
denote the secondary and primary masses), and the lack of normal
outbursts; thereby, a large amount of matter is expected to be stored
before superoutbursts (\cite{osa95wzsge}). Hence, early superhumps are
probably variations occurring in such a large disk, which definitely
shows strong tidal distortion. 

An important feature of early superhumps is a large variety of their
amplitudes; it is larger than 0.4~mag in some objects, while it is
smaller than 0.1~mag, or not significantly detected, in other objects
(e.g., \cite{kat02wzsgeESH}). In particular, edge-on (or a high
inclination) systems tend to exhibit large amplitudes of early
superhumps. This feature suggests that early superhumps can be attributed
to the rotation effect of a disk that has a non-axisymmetric
vertical structure (\cite{nog97alcom}). \citet{mae07bcuma} reproduce
the light curve of early superhumps in BC~UMa, assuming asymmetric
spiral-patterns in the disk.

\citet{osa02wzsgehump} propose that early superhumps are caused by
a two-armed spiral pattern that is generated by the 2:1 resonance. In
their model, the observed amplitude can be explained when the
geometrical effect of disks is considered, that is, the vertical
expansion along the spiral pattern. SPH simulations have shown the
appearance of the two-armed spiral pattern at the 2:1 resonance radius
(\cite{kun04esh}; \cite{kun05esh}). A non-axisymmetrically expanded
structure can be expected even without the resonance (\cite{sma01tidal};
\cite{ogi02tidal}). \citet{kat02wzsgeESH} propose that early
superhumps can be interpreted as irradiated emission of the elevated
surface of the disk caused by the vertical tidal deformation. In order
to evaluate those models, detailed observations have been long awaited
for early superhumps.

Most recently, \citet{mat09v455} (Paper~I) succeeded in measuring the
color variations associated with early superhumps for the first
time. In Paper~I, multi-band photometric observations during the
superoutburst of V455~And were reported. The observations revealed that
the hump component was redder than the stationary component. Paper~I
suggest that the source of early superhumps is a vertically-expanded
low-temperature region, probably located in an outer part of the
disk. In addition, it is interesting that the phase at which the
object was reddest significantly shifted to the phase of the flux
maximum. If the vertically-expanded region is limited at the outermost
region, the flux maximum coincides with the reddest phase. The 
observed phase shift implies that a part of the vertically-expanded
region is located not in the outermost part, but in a relatively
inner part of the disks. 

The idea of the reconstruction of the geometrical structure of disks
is then motivated by the color behavior of early superhumps in
V455~And. The phase of the humps would have information about the
azimuthal structure of the disk, while the color of the humps would 
have information about the radial structure. Therefore, the height
of each point in the disk could, in principle, be reconstructed using 
multi-band light curves of early superhumps. It is well known that
the intensity map of the disk in dwarf novae can be reconstructed by
using a tomography technique; the eclipse mapping uses the light curves of 
eclipses (\cite{hor85mapping}), and Doppler tomography uses
variations in emission-line profiles (\cite{mar88mapping}). In
addition to these methods, we propose a new tomography
method to reconstruct the height map of disks using early superhumps,
which we call the ``early superhump mapping''. 

In this paper, we present our model for early-superhump mapping
that is based on a Bayesian technique. The model is described in the
next section. We also present demonstrations of early superhump
mapping using artificial data sets in this section. In section~3, we
use this method for the light curves of V455~And. We discuss the
dependence based on assumed parameters in the model and the temporal
evolution of the reconstructed disk-structures.  In section~4, the effect
of irradiation is evaluated. We also discuss the physical implication
obtained from the reconstructed disk-structure. Finally, we summarize
our findings in section~5. 

\section{Model}

We consider a disk with a height map, $\{h(r,\theta)\}$, where $r$ and
$\theta$ is radial and azimuthal components in the cylindrical
coordinate system whose origin is located at the disk center;
$r=\sqrt{x^2+y^2}$ and $\tan \theta=y/x$ $(0\leq \theta <
2\pi)$. We take the $x$-axis in the direction of the center of the
secondary star, and the $y$-axis in the direction of the motion of the
secondary. The center of the secondary star is located at
$(x,y)=(1.0,0.0)$, and the binary separation is set to be
$a=1.0$. Hence, $h=h/a$ in this paper.

Our model assumes that all emission that we observe is that
from the disk surface defined with $\{h\}$. Then, the observed flux
density, $f_\nu(\phi)$, at a given frequency, $\nu$, and a given
phase, $\phi$ can be described as a function of the inclination
angle, $i$ and $\{h\}$. We develop a model to estimate $\{h\}$ from
$f_\nu(\phi)$, assuming $i$. The reconstruction of $\{h\}$ is not only an
inverse  problem, but also a non-linear problem, because a part of the
disk could be occulted by the other parts of the disk or by the
secondary star. We approach this problem with a Bayesian
framework. Our Bayesian model estimates the posterior distribution of
$\{h\}$ by using the Markov chain Monte Carlo (MCMC) method.

\subsection{Geometry and Emission from the Disk}

We calculate the flux from the disk at each phase using a similar
method reported in \citet{hac01RN}. The disk surface is divided into
small patches. A parch is defined as a closed region bounded by four
segments of grid lines. In the azimuth direction, the grids divide the
disk into $N_\theta$ equally-spaced bins. In the radial direction, the
disk is divided into $N_r$ bins between the inner radius, $r_{\rm in}$
and the outer radius, $r_{\rm  out}$. The width of the radial bin is
equally-spaced in a logarithmic scale; the $i$-th coordinate in $r$ is
described as $\log_{10} (r_i/r_{\rm in}) = \log_{10} (r_{\rm
  out}/r_{\rm in})\times i/N_r$. The center position, area, and normal
vector of each patch are defined as the same manner as described in
\citet{hac01RN}. 

Each patch emits blackbody radiation with a temperature of $T$ at the
center of the patch. We assume the following temperature distribution of the
standard accretion disk model (\cite{sha73disk});
\begin{eqnarray}
T = T_{\rm in} ({r\over r_{\rm in}})^{-3/4},
\end{eqnarray}
where $T_{\rm in}$ is the temperature at the innermost part of the
disk. The flux from each patch can be calculated from $T$, the size
and normal vector of the patch, and a viewing angle defined by $\phi$
and $i$. The flux in a given photometric band is calculated as the
flux density of the blackbody radiation at the center wavelength of
the band. We used $g$, $V$, $R_c$, $I_c$, and $J$ bands, and their
center wavelengths were set to be $0.4858$, $0.5505$, $0.6588$,
$0.8060$, and $1.2150$~${\rm \mu m}$, respectively.

The total flux at a given phase is a sum of the flux from all visible
patches in the viewing direction corresponding to the phase. We
consider the self-occultation in the disk; that is, the flux from the
patch is not counted if the center of a patch is occulted by the other
patch. In addition, the eclipse of the disk by the secondary star is 
also considered. The flux from a patch is counted if the patch center
is outside the Roche-lobe area projected perpendicular to the
viewing direction. We neglect the emission from the white dwarf (WD) and
boundary layer. They mainly contribute to the high energy emission
from an inner region, and less contribute to the emission from an
outer region where the early superhump originates. Thus, we calculate
multi-band light-curves, $f_\nu(\phi)$ from a set of
$h(r,\theta)$. The model light-curves are normalized by the average
flux for each band in order to be compared with the observed
light-curves. 

\subsection{Bayesian Model}

In general, a Bayesian approach provides a means for estimating the
posterior probability distribution of model parameters from the 
likelihood function and prior probability distribution. Our Bayesian
model estimates the posterior probability of $\{h(r,\theta)\}$,
$P(h)$. The model can be expressed as follows:
\begin{eqnarray}
P(h) = c L(f_{\nu,obs},f_{\nu,model}) \pi(h),
\end{eqnarray}
where $L$ is the likelihood function and $\pi$ is the prior
distribution. The normalization factor, $c$ is not important for
estimating $P(h)$ by using the MCMC method.

The likelihood function, $L$, is defined with the observed and model
light curves for each band. The model light curve is derived from a
given $\{h\}$, as described in the last subsection. The model assumes that
the observed fluxes have a Gaussian distribution with a mean of the
model fluxes and a variance of $\sigma^2$. Here, $\sigma$ corresponds
to photometric errors of the observed light curve. Hence, $L$ is
expressed as;
\begin{eqnarray}
L = \prod_{i,j} {1 \over \sqrt{2\pi \sigma_{i,j}^2}} \exp\left(-{[f_{\nu_i, obs}(\phi_j) - f_{\nu_i, model}(\phi_j)]^2 \over 2\sigma_{i,j}^2}\right),
\end{eqnarray}
where the subscripts $i$ and $j$ denote the band and phase,
respectively. 

The prior distribution, $\pi$, is a function of $\{h\}$, and essential to
solve the present inverse problem. The prior probability distribution
in our model consists of two components. The first one is for the
smoothness of the disk structure. It has a form: 
\begin{eqnarray}
\pi_{\rm smooth}&=& \nonumber \\
{1 \over \sqrt{2\pi w^2}} \prod_{l,m} &[&
\exp\left(-{[h_{l,m}-2h_{l-1,m}+h_{l-2,m}]^2 \over 2w^2}\right) \nonumber \\
&&\exp\left(-{[h_{l,m}-2h_{l,m-1}+h_{l,m-2}]^2 \over 2w^2}\right) ],
\end{eqnarray}
where $h_{l,m}$ denote the $l$-th and $m$-th grid points in
the radial and azimuthal directions, namely,
$h_{l,m}=h(r_l,\theta_m)$ ($l=1, \cdots, N_r+1$ and $m=1, \cdots, N_\theta$).
This prior distribution means that a sequence of the second difference
of $\{h_{l,m}\}$ in the radial and azimuthal directions follows a 
normal distribution having a standard deviation of $w$. A smoother
height-structure provides a higher $\pi_{\rm smooth}$. The weight
parameter $w$ plays a role in adjusting the smoothness of estimated
disk-structures. 

The second prior distribution provides a default map of $\{h_{l,m}\}$. It
is widely accepted that the emission of outbursting dwarf-novae can be 
explained by the standard-disk model (\cite{sha73disk}). In this
model, viscous heating in the disk is balanced by cooling by 
blackbody radiation from the disk surface. Its vertical structure is
determined by hydrostatic equilibrium. The disk height is expected to
be proportional to the radius in this scheme. In our model, we define, $h_{\rm
  disk}=0.1r$, and the prior distribution, $\pi_{\rm disk}$ as a
product of truncated normal distributions with means and standard
deviations of $h_{\rm disk}$: 
\begin{eqnarray}
&&\pi_{\rm disk} = \nonumber\\ 
&&\left\{ \begin{array}{ll} 
  \prod_{l,m} {1 \over \sqrt{2\pi h_{{\rm disk,}l,m}^2}} \exp\left(-{[h_{l,m}-h_{{\rm disk},l,m}]^2 \over 
       2h_{{\rm disk,}l,m}^2}\right) & (h_{l,m}\geq 0)\\ 
  0 & (h_{l,m}<0).
  \end{array} \right .
\end{eqnarray}
This prior distribution is useful to reject a height map containing
too large $h$, which violates the assumption of the standard-disk model
that the disk is geometrically thin. The assumed $h_{\rm disk}$ could
be much larger than those expected in the standard disk of
dwarf novae. According to \citet{kat08book}, the disk height is
estimated to be $h/r \sim c_s/r\Omega_K$, where $c_s$ and $\Omega_K$
denote the sound speed and Keplerian angular speed at distance $r$,
respectively. For typical parameters of dwarf novae ($T=10^4$~K,
$r_{\rm out}=10^{10}$~cm, and $M_{WD}=0.6M_\odot$), it gives $h\sim
0.01r$. This is one order smaller than that we assumed. However, we
consider accretion disks that could be deformed by a strong tidal
effect. In such a condition, the hydrostatic balance definitely
changes, and thereby the expected height-map could also
change. Therefore, $\pi_{\rm disk}$ defined in equation~(5) is a
reasonable choice as a default map. The prior distribution of our
model is, in total, expressed as $\pi$=$\pi_{\rm   smooth}$ $\pi_{\rm
  disk}$. 

Taken together, we estimate $P(h_{l,m})$ having $N_h =N_\theta
  (N_r+1)$ elements (including disk edges), using  $N_{\rm data}
=N_iN_j$ photometric data. The means of $P(h_{l,m})$ give 
$h_{l,m}$. We estimate $P(h_{l,m})$ using a MCMC algorithm. In
Appendix, we present a detailed description of 
our MCMC calculation. In the present work, we used $\sigma_{i,j}=0.01$~mag,
$N_\theta=20$, $N_r=16$, and $w=1.0$. A larger number of grids makes 
the amount of calculation significantly larger and the convergence
slower. The above values of $N_\theta$ and $N_r$ are the allowable
maximum ones in terms of the calculation time. The parameters $\sigma$
and $w$ correspond to the weights of the likelihood and prior
distribution components. A small value of $\sigma/w$ leads to a slow
convergence. In the present model, all of those parameters ($\sigma$,
$N_\theta$, $N_r$, and $w$) are related to a kind of ``resolution'' of
reconstructed images. Examples for different values of those
parameters are given in Appendix. 

\subsection{Demonstration with Artificial Data}

\begin{figure}
  \begin{center}
    \FigureFile(80mm,80mm){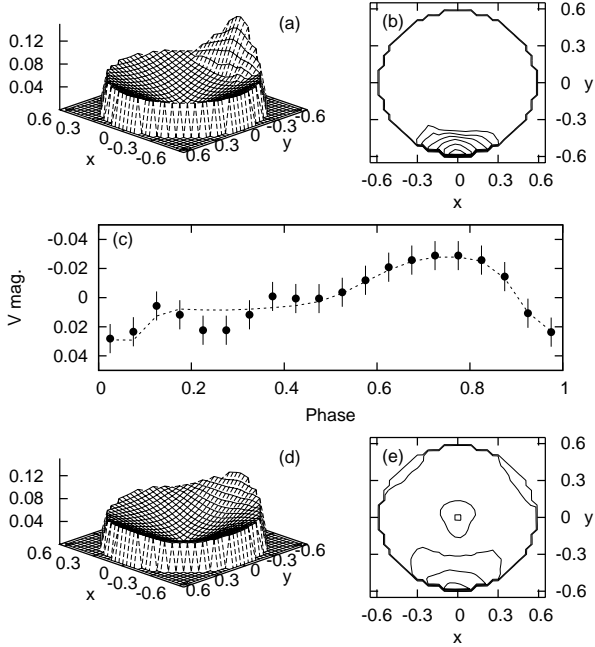}
  \end{center}
  \caption{Demonstration of the early superhump mapping for
    case~1. Panel~(a): assumed disk structure. Panel~(b): contour of
    the height ratio, $h/r$. Panel~(c): $V$-band light curve. The
    filled circles and dashed line represent the artificial data and
    model light curve, respectively. Panels~(d) and (e): the same as
    (a) and (b), but for the reconstructed disk structure. The center
    of the secondary star is located at $(x,y)=(1.0,0.0)$ in
    panels~(a), (b), (d), and (e). Panels~(b) and (d) show contours of
    $h/r=0.07$---$0.21$ with an interval of 0.02. The height maps are
    shown by linear interpolations of the original height grids,
    $\{h(r_l,\theta_m)\}$ $(l=1, \cdots, N_r+1$,
    $m=1, \cdots, N_\theta)$.}\label{fig:demo1} 
\end{figure}

\begin{figure}
  \begin{center}
    \FigureFile(80mm,80mm){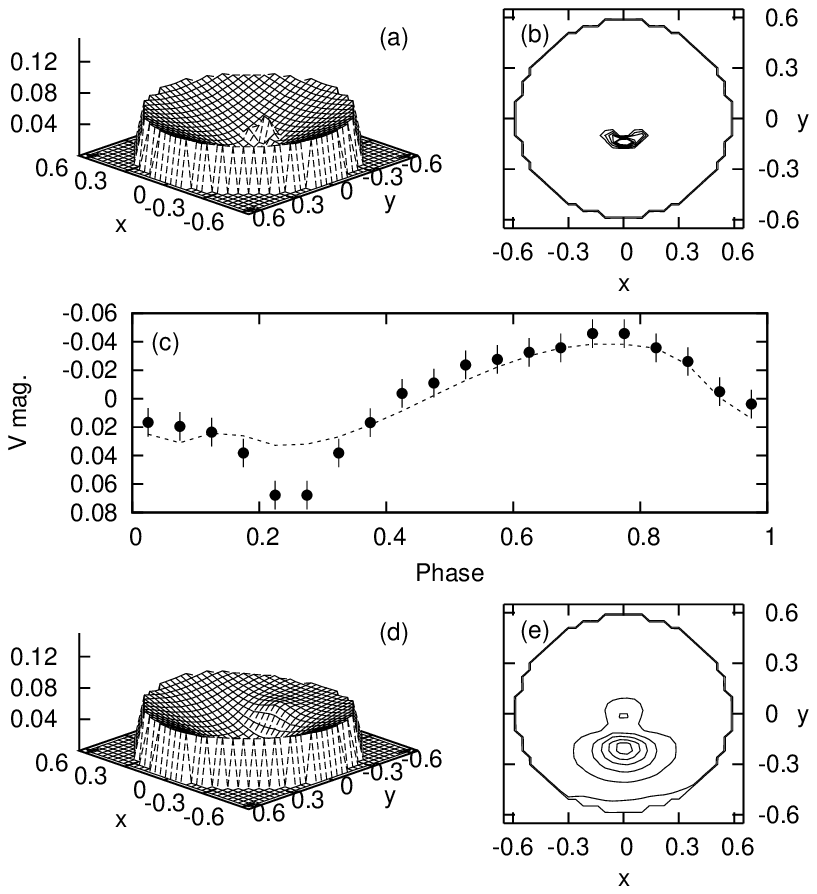}
  \end{center}
  \caption{The same as figure~\ref{fig:demo1}, but for case~2.}\label{fig:demo2}
\end{figure}

\begin{figure}
  \begin{center}
    \FigureFile(80mm,80mm){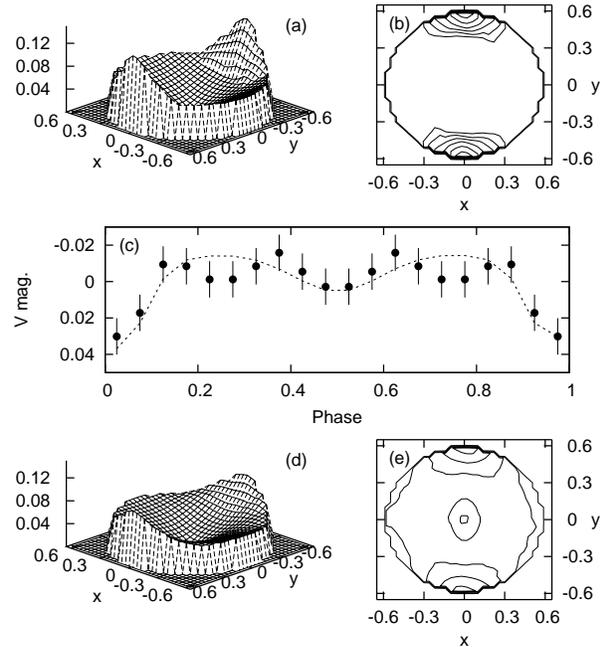}
  \end{center}
  \caption{The same as figure~\ref{fig:demo1}, but for case~3.}\label{fig:demo3}
\end{figure}

In this subsection, we present several demonstrations of our early
superhump mapping using artificial data sets. Figure~\ref{fig:demo1},
\ref{fig:demo2}, and \ref{fig:demo3} show three sets of artificial
data and results of the early superhump mapping, which we call 
cases 1, 2, and 3, respectively. In all three cases,
non-axisymmetrically flaring regions are assumed to be superimposed on
a disk having $h_{\rm disk}=0.1r$. In those figures, panel~(a) depicts
the assumed disk structure. Panel~(b) shows contours of the height
ratio, $h/r$. Non-axisymmetric features can be seen more conspicuously
in panel~(b). The light curves of the $g$, $V$, $R_c$, $I_c$, and
$J$-bands were calculated by ``observations'' of those disks, as
described in the last subsection. The $V$-band light curves are shown
in panel~(c), as indicated by the filled circles. We assumed a
photometric error of 0.01~mag in all bands and phases. The dashed
lines denote the model light curves. Phase~0 is defined as the time of
the mid-eclipse. Panels~(d) and (e) are the same as (a) and (b), but
for the reconstructed height map. The model parameters are listed in
table~\ref{tab:v455}. 
 
In case~1, the disk has a flaring part in the outermost region of the
disk. The peak height is 0.14 ($h/r=0.23$). In case~2, the disk has a
flaring part in a relatively inner region. The peak height is 0.05 at 
$r=0.14$ ($h/r=0.37$). As can be seen from figures~1 and 2, the
obtained height maps successfully reproduce the general profile of the
assumed disk structures. It should be emphasized that the model can
properly resolve the structures in the radial direction, namely in the
outer (case~1) and inner (case~2) structures, as well as the azimuthal
direction. In case~3, the disk has two flaring parts. Each of them has
the same structure as case~1. This is actually expected in
double-peaked early superhumps. The model again succeeded in
reconstructing the two-flaring structure, as can be seen in figure~3.

On the other hand, those demonstrations indicate that the early
superhump mapping could produce spurious signals in the estimated height
map. First, the peak heights are underestimated in all cases. The
peaks were assumed to be $h/r=0.23$ and $0.37$ in cases~1 and 2,
although they were $h/r=0.16$ and $0.23$ in the reconstructed maps,
respectively. In addition, the flaring areas were overestimated, or
smeared out. These results were obtained because the prior
distribution, $\pi_{\rm smooth}$ favors smooth structures. As a
result, the model light curves partly failed to reproduce the
substructures of the light curves; in figures~1 and 2, we can see dips
around phase 0.25 in the light curves, which are not reproduced by the
model. This dip is caused by occultation of the inner regions by the
assumed flaring parts. In the reconstructed height map, this flaring
part should be higher than the estimated one in order to reproduce the
dips. However, a disk with such a spiky structure reduces $\pi_{\rm
  smooth}$, and thereby reduces the posterior probability. Thus, our
model could fail to reproduce highly localized, or quite steep
height-structures in the disks. In principle, those smoothing effects
could be less significant if we apply a high $w$ in $\pi_{\rm
  smooth}$. A high $w$, however, leads to a slow convergence and
mixing in the MCMC procedure (for a detail, see the Appendix).
 
Second, in case~3, low $h$ regions appear in both the left and right
sides in the outermost parts in the reconstructed height map. The
height was assumed to be constant at  $h/r=0.10$ in the outermost
region, except for the flaring parts. The model estimated $h/r\sim
0.07$ in the left and right sides. As a result, the reconstructed disk
apparently has not a circular, but a box-like height-map. This is also
an artifact generated by $\pi_{\rm smooth}$. We should discuss
reconstructed height-maps while paying attention to this low $h$ pattern when
the double-peaked light curves are analyzed. 

Finally, we can see that the height of the innermost region is
slightly overestimated in all cases. They are estimated as $h/r \sim
0.12$. This is due to the prior distribution, $\pi_{\rm disk}$. Since
it is a product of truncated normal distributions, as defined in
equation~(5), its median is slightly larger than $h_{\rm disk}$. In
the artificial data, the light source is mainly in an outer region
of the disk, and the contribution of the innermost region is
relatively small. Hence, the height there almost follows $\pi_{\rm
  disk}$, and thereby slightly large $h$ are obtained.

Thus, we have demonstrated that our method can reconstruct the general
trend of accretion-disk structures, while reconstructed disks have
several noteworthy artifacts in their structure. In the next section,
we apply this model to real observations of early superhumps in
V455~And. The parameter dependence is also discussed in the next
section. 

\section{Results}
\subsection{Data and Model Parameters}

We have applied our Bayesian model to the data of V455~And reported
in Paper~I. V455~And is a WZ~Sge-type dwarf nova, whose
first-ever-recorded superoutburst was discovered in 2007
September. Their observation revealed color variations
associated with early superhumps for the first time (see
section~1). Paper~I reports simultaneous 6-band, $g$, $V$, $R_c$,
$I_c$, $J$, and $K_s$-band data on JD~2454354, which is the fifth day
(Day~5) since the outburst maximum. We used the blue five-band
observations on Day~5 for the present study, since Paper~I suggests
that the $K_s$-band flux needs another emitting source in addition to
the disk. High quality data not of all the five bands, but of two
bands ($V$ and $J$-band) were also available on JD~2454352 (Day~3). We
can study the temporal evolution of the disk structure using those two
sets of data.  

\begin{figure}
  \begin{center}
    \FigureFile(80mm,80mm){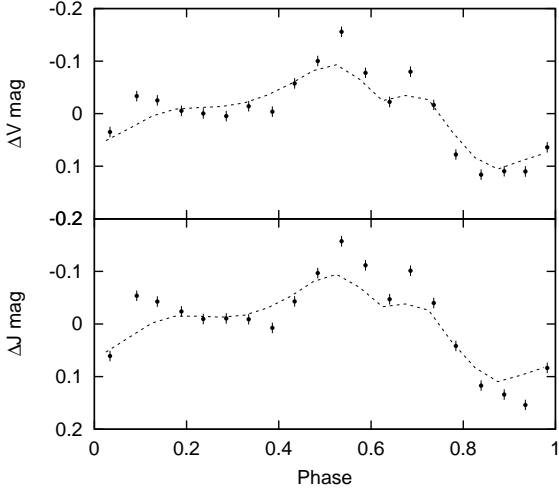}
  \end{center}
  \caption{Light curves of early superhumps of V455~And on Day~3. The
    upper and lower panels show $V$ and $J$-band light curves,
    respectively.  The filled circles represent the observed light
    curves. The small vertical bars indicate typical photometric
    errors of 0.01~mag. The dashed lines
    indicate the model light curves. The magnitudes are differential
    ones from the average magnitudes for each band.}\label{fig:v455d3lc} 
\end{figure}

\begin{figure}
  \begin{center}
    \FigureFile(80mm,80mm){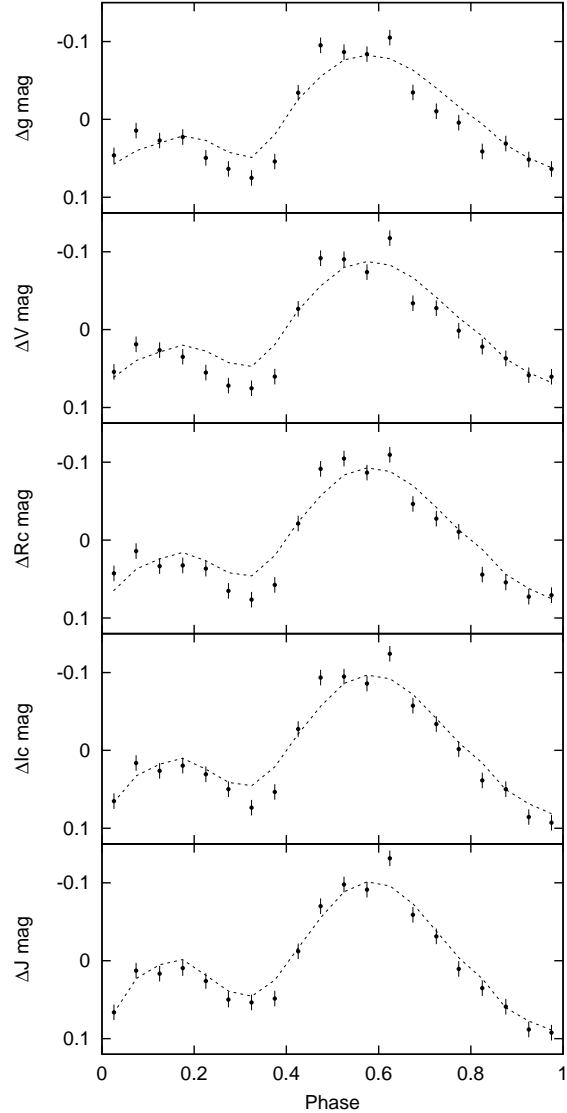}
  \end{center}
  \caption{Same as figure~\ref{fig:v455d3lc}, but for Day~5. From
    top to bottom, the panels show $g$, $V$, $R_c$, $I_c$, and
    $J$-band light curves.}\label{fig:v455d5lc} 
\end{figure}

The light curves were phase-averaged into 20 bins. Our model considers
eclipses of the disk by the secondary star. Hence, the epoch of
ephemeris is important for the conversion from time to phase, because
the timing of eclipses is sensitive to the orbital
phase. \citet{ara05v455} derived an ephemeris of V455~And using 
pre-outbust data between 2000 and 2003; ${\rm HJD} 2451812.67765(35) +
0.05630921(1)\times E$. Based on eclipse observations after the 2007
superoutburst (\cite{kat09pdot}), we found that the orbital phase of
the mid-eclipses is shifted by $\Delta \phi = -0.01$ from those
defined by the ephemeris in \citet{ara05v455}. It is quite likely that
the phase shift is simply due to accumulated errors in the
ephemeris. In the present study, we calculated the orbital phase with
the ephemeris presented by \citet{ara05v455}, and then added $+0.01$
to the phase. The phase-averaged light curves on Day~3 and 5 are shown
in figures~\ref{fig:v455d3lc} and \ref{fig:v455d5lc}, respectively.  

We can see several characteristic features of early superhumps in
figures~\ref{fig:v455d3lc} and \ref{fig:v455d5lc}. The light curves  
have a double-peaked profile; the secondary maxima are followed by the
primary maxima. The amplitude of the hump decreases with time, from
0.3~mag on Day~3 to 0.2~mag on Day~5. A noteworthy feature in the
Day~3 light curves is that the minima of early superhumps are slightly
shifted from the orbital phase $0.0$. This implies that the emitting
source has a highly asymmetric structure. It is also noteworthy that 
the secondary minimum was quite deep on Day~5. It is even deeper than
the primary minimum in the blue $g$ and $V$-bands. 

In the case of V455~And, intermittent short-term variations were
occasionally superimposed on early superhumps (see Paper~I). For
example, we can see a sharp spike at phase 0.62 in the light curve on
Day~5. The observed light curve on Day~5 is shown in figure~5 of
Paper~I. As can be seen from that figure, a temporary sharp spike
appeared just after a primary maxima not in all humps, but only in a
hump. Another spike-like feature can be seen also in the light curve
on Day~3 at phase 0.68. Their short time-scale implies that they were
originated from an inner high-temperature region, and have a different
mechanism (also see, \cite{dmi11wzsge}). It is difficult to reproduce
those spike-like features by our early superhump mapping, as mentioned
in subsection~2.3. However, it is not a serious problem in the present
study because we focus on stable and global structures of the disk. 

The model contains several parameters to be determined. They are
about the geometry of the binary and MCMC calculation. The
binary-parameters include the inclination angle, $i$, the outer and
inner radii, $r_{\rm out}$ and $r_{\rm in}$, the inner (or outer)
temperature, $T_{\rm in }$ (or $T_{\rm out}$), and mass ratio, $q (\equiv
M_2/M_1)$. \citet{ara05v455} reports that the inclination angle of
V455~And is $i\sim 75$~deg based on the presence of shallow eclipses
in its quiescent light curve. We used it in this study. We can
estimate $q$ using the empirical relationship between the
superhump-period excess and $q$. According to \citet{kat09pdot}, the
superhump period excess of V455~And is $\varepsilon = (P_{\rm
  SH}-P_{\rm orb})/P_{\rm  orb}=0.015$, and the relation between
$\varepsilon$ and $q$ is $\varepsilon = 0.16q+0.25q^2$. Hence, the
mass ratio of V455~And can be estimated as $q=0.083$. Assuming a white
dwarf having a mass of $M_{\rm WD}=0.6 M_\odot$ (\cite{ara05v455}) and
a radius of $R_{\rm WD}=0.012 R_\odot$ (\cite{pro98WD}), we can then
calculate the binary separation to be $a=3.7\times 10^{10}\,{\rm
  cm}$. The model calculation was performed with $r_{\rm in}=R_{\rm
  WD}$ and $r_{\rm out}=0.6a$. This $r_{\rm out}$ condition is
expected from the 2:1 resonance radius (\cite{osa02wzsgehump}). 

The innermost disk temperature, $T_{\rm in}$, can be estimated
from observed colors of V455~And. They were $g-V=0.015$, $V-R_c=0.052$,
$R_c-I_c=0.039$, and $I_c-J=-0.090$ on Day~5 (Paper~I). These are
compared with the model color of an axisymmetric disk with $h=h_{\rm
  disk}$ and $T_{\rm in}$. The best-fitted temperature is $T_{\rm in}
= 8.2\times 10^4 \,{\rm K}$ (or $T_{\rm out} = 6.9\times 10^3 \,{\rm
  K}$) for the Day~5 data. In the same manner, the $V-J$ color of
$-0.034$ on Day~3 yields $T_{\rm in} = 8.6\times 10^4 \,{\rm K}$ (or
$T_{\rm out} = 7.3\times 10^3 \,{\rm K}$). 

We summarize the model parameters for the analysis of the V455~And
data in table~\ref{tab:v455}. In the next subsection, we report on the
reconstructed height maps calculated with those parameters for the
Day~5 data. In subsection~3.3, we discuss the parameter dependence of
the result by comparing the reconstructed height maps obtained with
different sets of $T$, $r_{\rm out}$, $i$, and $q$.

\begin{table*}
  \caption{Model parameters}\label{tab:v455}
  \begin{center}
    \begin{tabular}{lrrrrrr}
      \hline
      Case & $r_{\rm out}/a$ & $r_{\rm in}/a$ & $T_{\rm in}$ ($\times 10^4$ K) & $T_{\rm out}$ ($\times 10^3$K)& $i$ (deg) & $q$ \\
      \hline
      1,2,3 (subsection~2.3)   & 0.60 & 0.020 & 8.7 & 7.3 & 75.0 &  0.11\\
      \hline
      Day 5 (default) & 0.60 & 0.022 & 8.2 & 6.9 & 75.0 & 0.083\\
      Day 3 (default) & 0.60 & 0.022 & 8.6 & 7.3 & 75.0 & 0.083\\
      \hline 
      A       & 0.60 & 0.022 & 8.0 & 6.8 & 70.0 & 0.083\\
      B       & 0.60 & 0.022 & 8.7 & 7.3 & 80.0 & 0.083\\
      C       & 0.60 & 0.020 & 8.9 & 7.5 & 75.0 & 0.11\\
      D       & 0.60 & 0.022 & 8.2 & 6.9 & 75.0 & 0.067\\
      E       & 0.60 & 0.022 & 7.2 & 6.1 & 75.0 & 0.083\\
      F       & 0.60 & 0.022 & 9.2 & 7.8 & 75.0 & 0.083\\
      G       & 0.70 & 0.022 & 9.2 & 6.9 & 75.0 & 0.083\\
      H       & 0.50 & 0.022 & 7.3 & 7.0 & 75.0 & 0.083\\
      \hline
      Day 5 (irradiated, $T_{\rm cs}=1.2\times 10^5$~K) & 0.60 & 0.022
        & 6.0 & 5.8(min), 13.4(max) & 75.0 & 0.083\\
      Day 5 (irradiated, $T_{\rm cs}=2.0\times 10^5$~K) & 0.60 & 0.022
        & 6.0 & 5.8(min), 22.1(max) & 75.0 & 0.083\\
      \hline
    \end{tabular}
  \end{center}
\end{table*}

\subsection{Early Superhump Mapping for the Day~5 Data of V455~And}

\begin{figure}
  \begin{center}
    \FigureFile(80mm,80mm){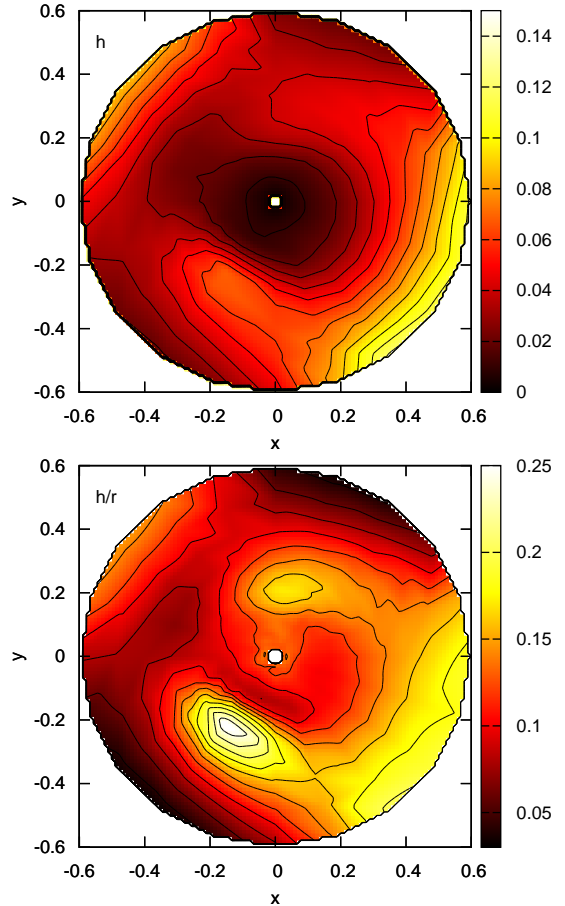}
  \end{center}
  \caption{Height map for the data on Day~5 of V455~And. The upper and
  lower panels show maps of $h(=h/a)$ and $h/r$,
  respectively. Both the contours and color-maps represent the same
  maps. The upper panel shows contours of $h=0.00$---$0.15$ with an
  interval of 0.01. The lower panel shows contours of
  $h/r=0.00$---$0.25$ with an interval of 0.02. The secondary star is
  located at $(x,y)=(1.0,0.0)$.}\label{fig:v455d5map}
\end{figure}

We estimated $\{h(r,\theta)\}$ using the data on Day~5 of V455~And with the
model parameters discussed in the last subsection. The estimated
$\{h(r,\theta)\}$ are represented by the height maps shown in
figure~\ref{fig:v455d5map}. The upper and lower panels 
show the height maps of $h(=h/a)$ and $h/r$, respectively. The model
light-curves calculated with this $\{h(r,\theta)\}$ are indicated by the
dashed lines in figure~\ref{fig:v455d5lc}. We can see that the observed
light-curves are well reproduced by the model. It apparently
fails to reproduce short-term variations, such as a spike-like feature
at phase~0.62. This can be explained by the fact that $\pi_{\rm smooth}$
favors smooth structures, as mentioned in section~2 and also in the
last subsection. The secondary minima of the model light curves are
slightly shallower than the observed ones, particularly in the blue
bands. 

The maximum $h$ region is distributed in the outermost part of
the right-lower quadrant of the map, as shown in
figure~\ref{fig:v455d5map}. This part is responsible for the primary
maximum in the light curve. The highest point in the disk is $h_{\rm
  max}=0.12$ at an outermost point of $(x,y)=(+0.36,-0.48)$
($h/r=0.20$). The elevation angle of this point from the disk center
is, $\arctan (h_{\rm max}/r_{\rm out}) = \arctan (0.12/0.60) =
11\,{\rm deg}$. This means that the center of the disk is occulted by
this outermost flaring part in the case of $i\gtrsim 79$~deg. The
secondary maximum in the light curve is generated by a flaring
outermost part in the left-upper quadrant of the map. The maximum
$h$ in this part is $h=0.09$ at $(-0.49,+0.34)$ ($h/r=0.15$). This is
smaller than $h_{\rm max}$. Hence, the difference between the
amplitudes of the primary and secondary maxima can be explained by the
difference between the heights of the corresponding flaring parts. 

In addition to those two major features, we can see flaring arm-like
patterns elongated to relatively inner parts of the disk. They can be
seen more conspicuously in the lower panel of
figure~\ref{fig:v455d5map}. In the left-lower quadrant, there is a
``spot''-like high region, which could be connected to the
primary maximum part in the right-lower quadrant. This region includes
the largest $h/r$ point among the whole disk, $h/r=0.26$ at
$(-0.16,-0.21)$. The observed deep secondary minimum in the light
curves is generated by the occultation of inner regions by this 
flaring area. The relatively shallow secondary minimum of the model
light curves implies that the flaring part would originally have a
more localized structure with larger $h$ than the reconstructed
disk. The other arm-like pattern can also be seen in the right-upper
quadrant. Its $h/r$ are relatively moderate ($h/r \sim 0.19$) compared
with those of the left-lower arm.

There are low regions in the outermost parts of both the left-lower
and right-upper quadrants. This could be partly due to the spurious
signal, as mentioned in subsection~2.3. However, the height in 
those regions is even smaller than that of case~3 in subsection~2.3
in which $h/r \sim 0.07$; in the case of figure~\ref{fig:v455d5map},
the minimum $h$ point at $(-0.34,-0.49)$ has $h/r=0.03$. It indicates
that those low regions are not only a totally spurious signal, but also
an intrinsic feature of the disk. 

\subsection{Dependence on the Model Parameters}

\begin{figure*}
  \begin{center}
    \FigureFile(165mm,165mm){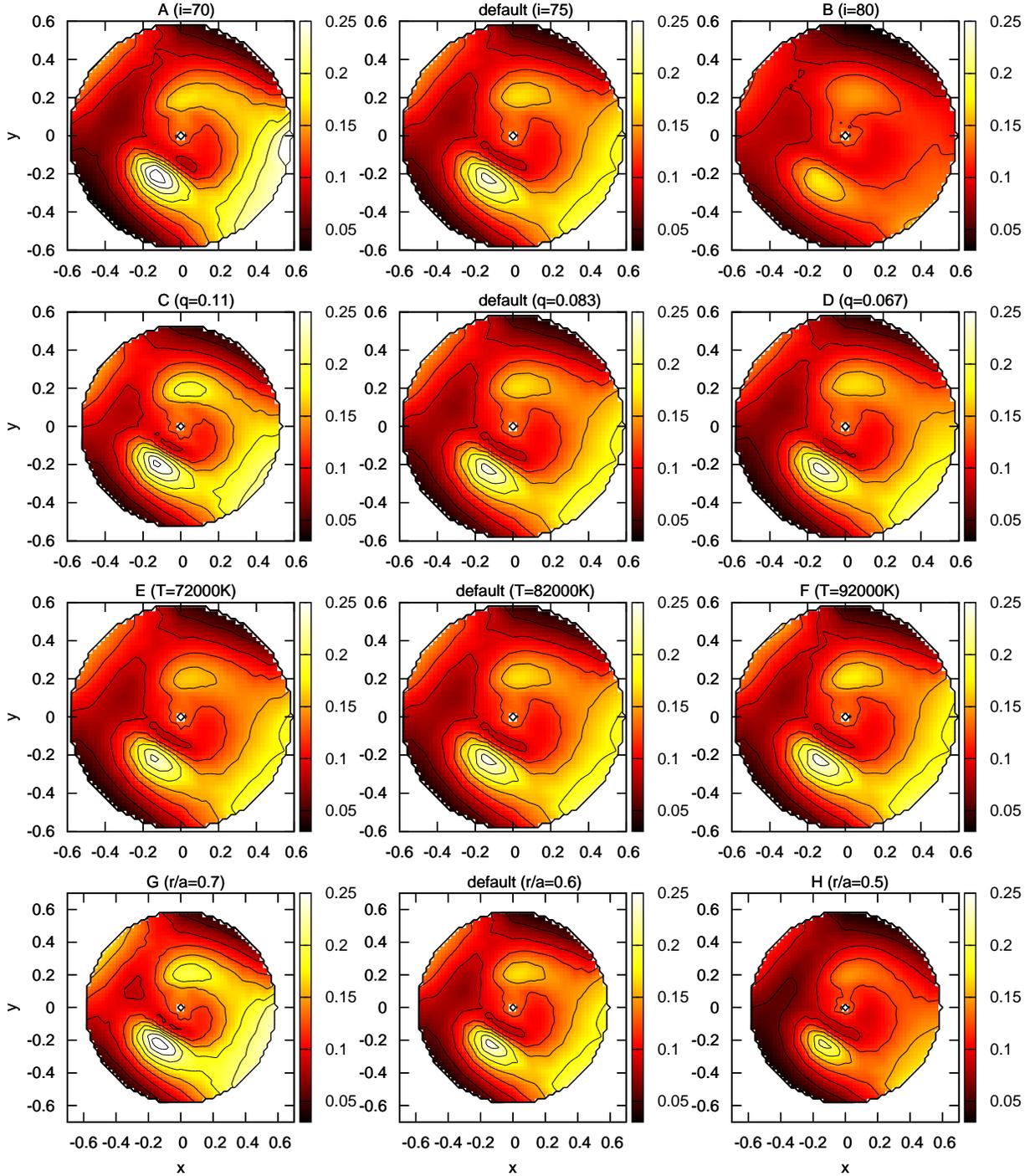}
  \end{center}
  \caption{Height maps calculated from the data on Day~5 of V455~And
    for different parameters. The contours are shown from
    $h/r=0.00$---$0.27$ with an interval of 0.03. The secondary star is 
    located at $(x,y)=(1.0,0.0)$. The model parameters for each case
    are listed in table~\ref{tab:v455}. From top to bottom, they
    were calculated with different $i$, $q$, $T_{\rm in}$, and $r_{\rm
      out}$, respectively. All middle-column panels are the same as
    figure~\ref{fig:v455d5map}, shown for a comparison.}\label{fig:parms_map}  
\end{figure*}

\begin{figure*}
  \begin{center}
    \FigureFile(165mm,165mm){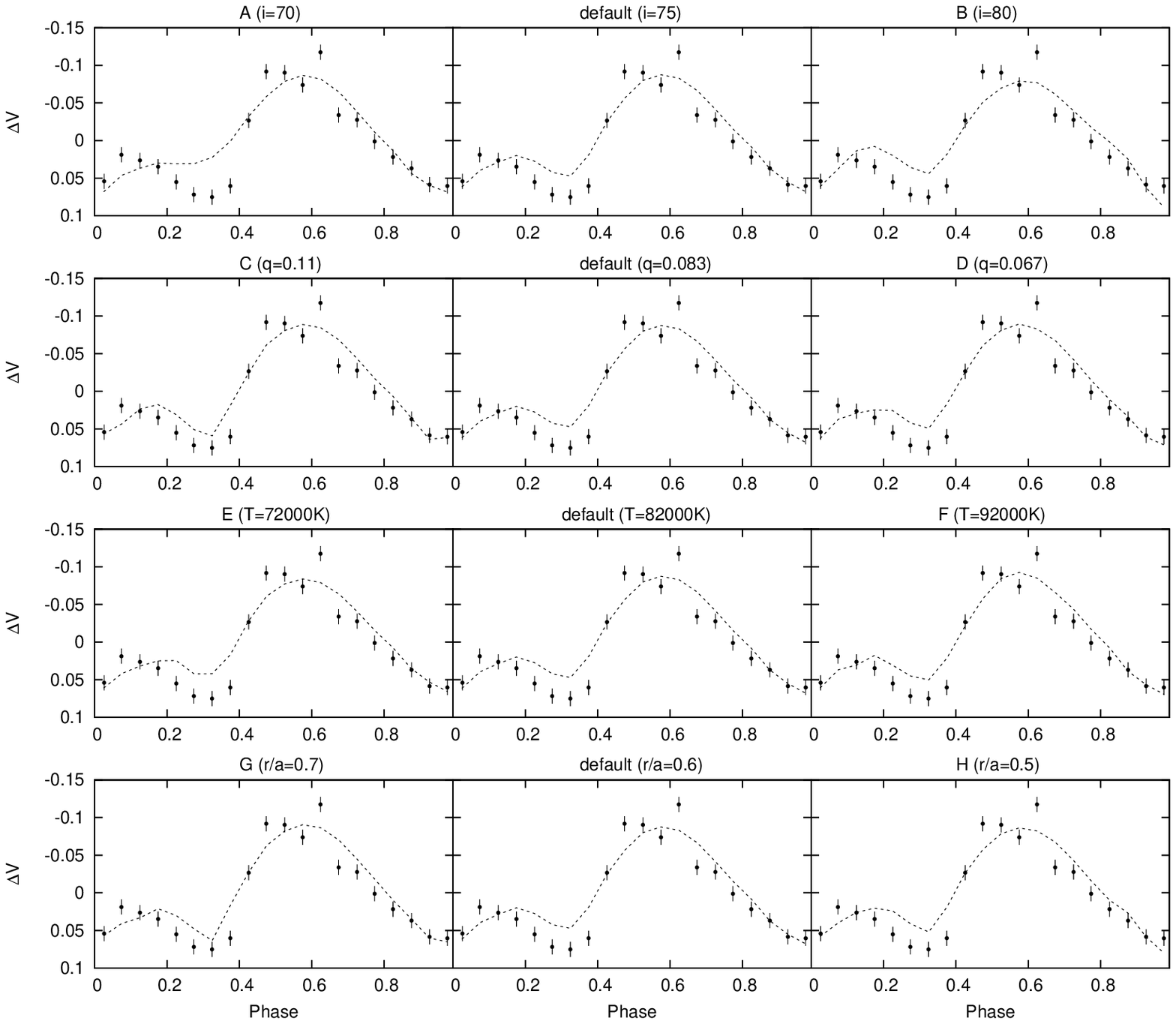}
  \end{center}
  \caption{Model light curves calculated from the data on Day~5 of
    V455~And for different parameters, indicated by the dashed
    lines. The filled circles indicate the observed light curves. The
    model parameters for each case are summarized in
    table~\ref{tab:v455}. From top to bottom, they were calculated
    with different parameters $i$, $q$, $T_{\rm in}$, and $r_{\rm
      out}$. All middle-column panels are the same as
    figure~\ref{fig:v455d5lc}, shown for comparison.}\label{fig:parms_lc}
\end{figure*}

In this subsection, we examine the dependence of the result reported in 
the last subsection on the assumed parameters. We recalculated the height
maps for eight sets of different parameters of $i$, $q$, $T_{\rm in}$,
and $r_{\rm out}$ listed in table~\ref{tab:v455}, as labeled from
cases~A to H. Their height maps and model light-curves are depicted in
figure~\ref{fig:parms_map} and \ref{fig:parms_lc}, respectively. In
both figures, the middle-column panels show the same maps and light-curves as
reported in the last subsection. 

The dependence of the result on $i$ can be seen in cases~A and B, in
which we set $i=70.0$ and $80.0$~deg, respectively. These changes in $i$
required us to readjust $T_{\rm in}$ to be consistent with the observed
colors. We calculated $T_{\rm in}=8.0\times 10^4$~K for case~A and
$8.7\times 10^4$~K for
case~B. The other parameters are the same as those of the default
ones. As can be seen in figure~\ref{fig:parms_map}, the global pattern
of the height map is apparently insensitive to $i$, while there are
minor differences in the maps with different $i$. First, the disk
height apparently decreases as the increase in $i$. This can be
explained as follows: the amplitude of the humps can be considered as
a ratio of the projected area of high $h$ region to low $h$ region on
the plane perpendicular to the viewing direction. With a given height
map, the area ratio decreases with a decrease in $i$. Second, the model
light curve of case~A fails to reproduce the deep secondary minimum,
as can be seen in figure~\ref{fig:parms_lc}. In a smaller $i$ case,
like case~A, the disk is required to have higher $h$ to explain the
observed amplitude. However, quite high $h$ values reduce the posterior
probability, because $\pi_{\rm disk}$ becomes small. Hence, the
reconstructed map with a small $i$ fails to reproduce substructures in
the light curve.

Another noteworthy feature of the height map is a high region around 
$(x,y)=(0.6,0.0)$, which is unnaturally confined to a small region. It 
is most prominent in case~A, and less prominent in higher $i$
cases. This high region means that the inner region of the disk is
required to be occulted at phase~$\sim 0.0$ to explain the
observation. In a high-$i$ case (case~B), the eclipse by the secondary
star plays a role in hiding the inner region, and thereby the
high region around $(x,y)=(0.6,0.0)$ disappears. It implies $i\gtrsim
75$~deg for V455~And in order to avoid making this unnatural
feature. 

The mass ratio, $q$, is one of the most uncertain parameters, while
the reconstructed height maps are insensitive to small changes in $q$,
as shown in cases C and D in figure~\ref{fig:parms_map} and
\ref{fig:parms_lc}. The height maps and light curves of cases~C and D
were calculated with $q=0.11$ and $0.067$, respectively. In case~C, a
large $q$ slightly changes the binary separation, the inner radius,
and as a result, the inner temperature, as shown in
table~\ref{tab:v455}. The changes in those parameters are negligible
in case~D. The mass-ratio of case~C corresponds to
$\varepsilon=0.021$, which is the largest $\varepsilon$ among WZ~Sge
stars having very short $P_{\rm orb}$ of $<0.057$~d
(\cite{kat09pdot}). The mass-ratio only has a minor effect on the
results, because it is only related to the eclipse by the secondary
star. The mass-ratio is so small, even in case~C, that the small
changes in $q$ only lead to small changes in the Roche lobe
geometry. Hence, the uncertainty in $q$ has no significant effect on
the results in the case of WZ~Sge stars.  

The innermost temperature, $T_{\rm in}$, was estimated from the observed
color. While we use $T_{\rm in}=8.2\times 10^4$~K in the last
subsection, it has a 1-$\sigma$ error of $0.4\times 10^4$~K. Hence, we
should check the dependency of the result on $T_{\rm in}$. In cases~E
and F, we estimate $\{h\}$ with $T_{\rm in}=7.2$ and $9.2\times 10^4$~K,
which correspond to $T_{\rm out}=6.1\times 10^3$ and $7.8\times
10^3$~K, respectively. The reconstructed height-maps are consistent
each other, as can be seen from figure~~\ref{fig:parms_map}. Thus, the
reasonable changes in temperature have no significant effect on the
results of the early superhump mapping.

Next, we evaluate the dependency of the results on the outer
disk-radius, $r_{\rm out}$. We used $r_{\rm out}/a=0.6$ in the last
subsection, which is expected from the 2:1 resonance radius
(\cite{osa02wzsgehump}). It has, however, not been confirmed by
observations. In cases~G and H, the height maps were calculated with
$r_{\rm out}/a=0.7$ and $0.5$, respectively. According to
\citet{osa02wzsgehump}, the disk radius of case~G is close to the 3:1
resonance radius in WZ~Sge stars. The disk radius of case~H
corresponds to the maximum radius allowed by the tidal truncation. It
is noteworthy that the $h/r$ maps are quite similar in all cases with
$r_{\rm out}/a=0.5$, $0.6$, and $0.7$. As a result, $h$ is 
larger in the case with larger $r_{\rm out}$.

Thus, the results of the early superhump mapping are insensitive to
reasonable changes in the model parameters, $i$, $q$, $T_{\rm in}$,
and $r_{\rm out}$. The inclination angle has a relatively large effect
on the result. 

\begin{figure}
  \begin{center}
    \FigureFile(80mm,80mm){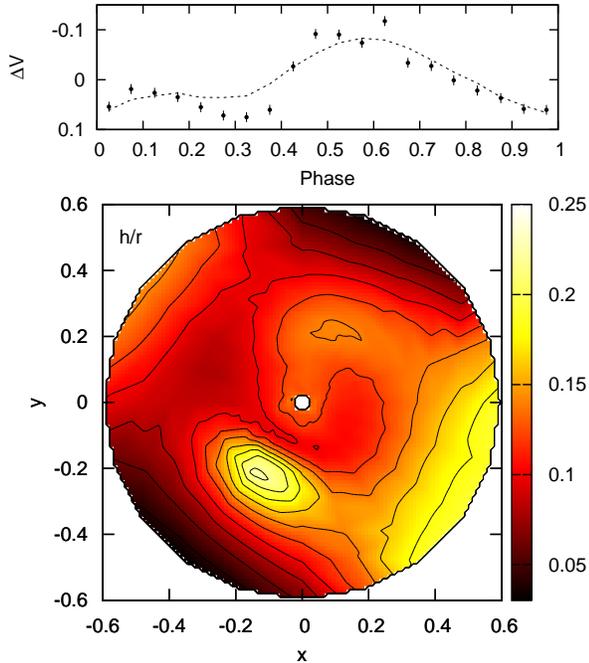}
  \end{center}
  \caption{Light curve (upper) and height map (lower) calculated from
    the $V$ and $J$-band light curves on Day~5 of V455~And. In the upper
    panel, the filled circles represent the observed light curves. The
    dashed line indicates the model light curve. The color maps and
    contours represent $h/r$. The contours are shown from 
    $h/r=0.00$---$0.25$ with an interval of 0.02, the same as
    figure~\ref{fig:v455d5map}. The secondary star is
    located at $(x,y)=(1.0,0.0)$.}\label{fig:v455d5VJ}  
\end{figure}

Finally, we discuss the dependence of the reconstructed height map on
the number of data. We evaluate it by comparing the height map
calculated with two-band data with that with five-band
data. Figure~\ref{fig:v455d5VJ} shows the light curve (upper panel)
and height map (lower panel), which were calculated from $V$- and
$J$-band data of V455~And on Day~5, instead of all five-band
data. Compared with figures~\ref{fig:v455d5lc} and
\ref{fig:v455d5map}, the result obtained with the two-band data
reproduces the major features of the height map obtained with the
five-band data: the two outermost flaring regions and two inner
arm-like patterns. The maximum $h$ in the height map calculated from
the two-band data is only 6\% smaller than that from the five-band
data. The structures in both the light curve and height map are
smoother than those obtained from the five-band data. This is 
due to the small amount of data, which makes the weight of the
likelihood function small with respect to the prior probability. 

Thus, we can obtain reconstructed height maps with a reasonable degree 
of accuracy, even in the case that only two-band data is
available. High quality data were obtained on Day~3 in Paper~I, but
only in the $V$ and $J$-bands. In the next subsection, we report the
results  obtained from the data on Day~3.

\subsection{Early Superhump Mapping for the Day~3 Data of V455~And}

\begin{figure}
  \begin{center}
    \FigureFile(80mm,80mm){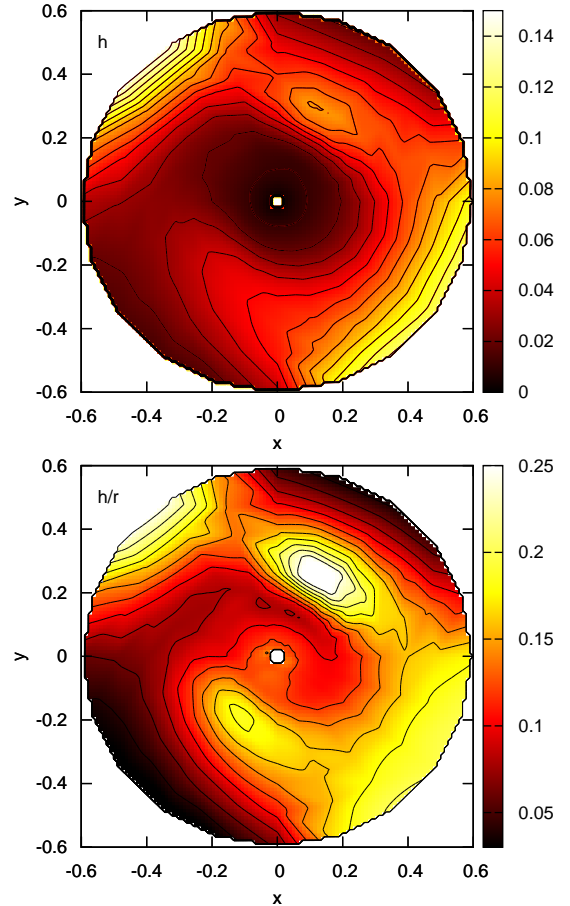}
  \end{center}
  \caption{Same as figure~\ref{fig:v455d5map}, but for
    Day~3.}\label{fig:v455d3map} 
\end{figure}

We estimated $\{h(r,\theta)\}$ using the $V$ and $J$-band data on Day~3 of
V455~And. The model parameters are listed in table~\ref{tab:v455}. The
reconstructed height map on Day~3 is shown in
figure~\ref{fig:v455d3map}. The model light-curve is indicated by the
dashed line in figure~\ref{fig:v455d3lc}. The height map on Day~3 has
common features with that on Day~5: the outermost flaring parts for
the primary and secondary maxima of the light curve and the inner
arm-like patterns.

On the other hand, there are several differences in the
detailed structures between them, which represent a temporal evolution of
the disk. The maximum height in the outermost part is $h/r \sim 0.25$ on
Day~3, while it is $h/r\sim 0.20$ on Day~5. The amplitude of the light
curve on Day~3 is larger than that on Day~5. The large amplitude on
Day~3 is, therefore, due to the large $h$. We confirmed that the disk
temperature has only minor effect on the amplitude; the amplitude
keeps small with $\sim 0.2$~mag in the light curve calculated with the
height map on Day~5 and the temperature on Day~3. In general, early
superhumps decrease in amplitude with time. Our result suggests that
it is due to the decrease in the disk height in the outermost region,
rather than a decrease of the disk temperature.

Another noteworthy feature is the arm patterns. In
figure~\ref{fig:v455d3map}, the right-upper arm has the largest 
$h/r$ on Day~3, while the left-lower arm has the largest $h/r$ on 
Day~5. The weakening of the left-lower arm is due to the shallow
secondary minimum in the light curve on Day~3. No highly flaring
region is required to reproduce such a shallow secondary minimum. 
The strong right-upper arm is required to reproduce the small dip at 
phase $\sim 0.63$ in the light curve. This dip feature may, however,
be related to the short flare-like feature around phase $\sim
0.68$; the small dip feature could be a spurious signal generated by the
intermittent short flare. The height of the right-upper arm could be
overestimated if the flare-like feature around phase $\sim 0.68$ has a
different origin from early superhumps.

\section{Discussion}
\subsection{Implication from the Reconstructed Disks}

Here, we compare the reconstructed disk-structures with theoretical
ones which have been proposed for the origin of early superhumps. As
mentioned in section~1, \citet{kat02wzsgeESH} propose the tidal
distortion scenario, while \citet{osa02wzsgehump} propose the 2:1
resonance scenario. 

\citet{ogi02tidal} calculate the disk structure which is tidally
distorted. They present the height map of a disk in a binary
system with $q=0.5$ in their figure~2. The tidally distorted disk also
has two flaring parts in its outermost region. We find that the phases of the
outermost flaring parts are consistent with those of our reconstructed
map. In addition, the inner arm-pattern shown in their figure is
reminiscent of the arm-pattern in the right-upper quadrant of the
reconstructed image in figure~\ref{fig:v455d5map}. Thus, the global
structure of the reconstructed image is similar to that expected from
the tidal distortion. 

However, the tidal distortion scenario cannot explain the arm-like
pattern in the left-lower quadrant of the reconstructed image. This
pattern is responsible for the secondary minimum in the light curve. 
As mentioned in subsection~3.2, that area has a quite large $h/r$, and
possibly has a spot-like structure, rather than arm-like one. In
conjunction with the height map on Day~3, this feature probably
evolved from Day~3 to 5. Two scenarios can be considered to
explain this flaring area. First, an additional effect may be
responsible for it. The tidal effect may affect the global
structure of the disk, while it is difficult to make such a
well-localized structure. It could be related to the accretion stream
from the secondary star or the evolving ordinary superhump. However,
these scenarios are highly speculative. Second, it is possible that
the inner arm-like pattern is a spurious structure if the model
assumptions are not valid for the real structure of the disks. In this
case, the most controversial assumption is the temperature
distribution of the disk, namely $T= T_{\rm in} (r/r_{\rm
  in})^{-3/4}$. If the temperature distribution is not axisymmetric,
it could make a height map without the inner flaring region. The
irradiation effect may be able to achieve such a condition. We
evaluate it in the next subsection. 

We should note that the height distortion calculated by
\citet{ogi02tidal} is for a binary system of $q=0.5$, which is much 
larger than those of WZ~Sge stars ($q\lesssim 0.1$). Detailed studies
should be carried out with theoretical disk-structure calculated with
appropriate mass-ratios in the future. \citet{kun04esh} present SPH
simulations of disks which extend to the 2:1 resonance radius (also see,
\cite{kun05esh}). Their result shows a two-armed density wave close to
the outermost region in the disk. The phases of this two-armed pattern
are consistent with the outermost flaring regions of our reconstructed
map. However, it is difficult to compare the result of their
SPH simulation directly with our height map. A study of the height
distortion expected from the SPH simulations is also required in
future.

\subsection{Effect of Irradiation}

We evaluate the effect of irradiation to the early superhump mapping
in this subsection. The temperature in the outermost flaring parts
would increase by the irradiation effect. If this is the case, the
temperature distribution could be non-axisymmetric, which may change
the structure of the reconstructed height maps. The irradiation effect
depends on the luminosity and geometry of the source for irradiation
and the albedo of the irradiated surface. Unfortunately, those
parameters are poorly known. In outbursting dwarf novae, the most
powerful source for irradiation is supposed to be boundary layers,
which are formed between the WD surface and the disk. X-ray and UV
observations suggest that its temperature is typically on the order of
$10^5$~K, while its size is poorly known (e.g., \cite{lon96ugem};
\cite{vanderwoe87CVXray}; \cite{mau00oycarEUVE}). The temperature and
geometry of boundary layers definitely depend on the accretion rate,
which can change during outbursts. However, no estimation of the
temperature and size of the boundary layer has been reported for
V455~And. Furthermore, V455~And is known as an intermediate polar
(\cite{ara05v455}). Its physical condition in the innermost region of
the disk could be different from that of non-magnetic dwarf novae in
outburst. Here, we focus on how the irradiation effect changes the
disk structures estimated by the early superhump mapping. Hence, we
consider an extreme condition that the total luminosity, $L_{\rm tot}$,
is dominated not by the luminosity generated by the viscous heating,
$L_{\rm vis}$, but by the irradiation luminosity, $L_{\rm irr}$, in
the outermost part. 

We calculated the irradiation effect using a similar method reported
in \citet{hac01RN}; a patch on the disk surface is irradiated by the
central source and other patches if there is no blocking patch
between the centers of the source and patch. The central source
for irradiation, which is probably dominated by the boundary layer
emission, has a luminosity of $L_{\rm cs}=4\pi r_{\rm in}^2 \sigma
T_{\rm cs}^4$. We assume that the radiation from the central source is
isotropic. Consider a patch, $p_i=p(r,\theta)$, whose size and normal
vector are $s_i$ and $\mathbf{n}_i$. The patch $p_i$
receives energy from the central source by irradiation, and releases
the energy by radiation with a luminosity $L_{\rm irr,cs}$, which is
calculated as
\begin{eqnarray}
L_{\rm irr,cs} = \eta \xi_{ci} \frac{L_{\rm cs}}{4\pi r^2} 
s_i (\mathbf{n}_i \cdot\mathbf{\hat{r}}_{ic}) ,
\end{eqnarray}
where $\eta$ is the fraction of the absorbed energy to the total
irradiated energy in $p_i$. $\xi_{ci}$ is $1$ if there is no blocking
patch between the centers of the source (the disk center in this case)
and the patch. Otherwise, $\xi_{ci}=0$. $\mathbf{\hat{r}}_{ic}$
denotes the unit vector in the direction of the patch center from the
disk center. Using a similar notation, the irradiation luminosity by
other patches is
\begin{eqnarray}
L_{\rm irr,disk} = \eta \sum_{j\neq i} \xi_{ij} \frac{\sigma T_j^4}{2\pi r_{ij}^2} 
s_j (\mathbf{n}_j \cdot\mathbf{\hat{r}}_{ji})
s_i (\mathbf{n}_i \cdot\mathbf{\hat{r}}_{ij}).
\end{eqnarray}
The total luminosity of $p_i$ is, hence, 
\begin{eqnarray}
\sigma T_{\rm irr}^4 s_i = L_{\rm tot}=L_{\rm vis}+L_{\rm irr,cs}+L_{\rm irr,disk}.
\end{eqnarray}
The original luminosity, $L_{\rm vis}$, is calculated as $L_{\rm
  vis}=\sigma T_{\rm org}^4 s_i$, where $T_{\rm org}$ is a temperature
given by an assumed temperature distribution. The temperature of the
irradiated patch, $T_{\rm irr}$, is given by equation~(8).

The structure of irradiated disks was estimated by iteratively using
our method of the early superhump mapping described in section~2. In
the $n$-th iteration, we calculate a height map, $\{h\}_n$ with an
old temperature distribution. Then, a new temperature distribution is
defined by $\{T_{\rm irr}\}_{n+1}$ which is given by the calculation
described above. We calculate a new height map, $\{h\}_{n+1}$ using 
the new temperature distribution. This process is repeated until
$\{h\}_n$ converge. The analysis of the data of V455~And requires
4---5 iterations for convergence. The initial temperature distribution
is given by a standard one, that is, $T=T_{\rm in}(r/r_{\rm
  in})^{-3/4}$, while we confirmed that the final result is
insensitive to the initial distribution.

\begin{figure*}
  \begin{center}
    \FigureFile(160mm,80mm){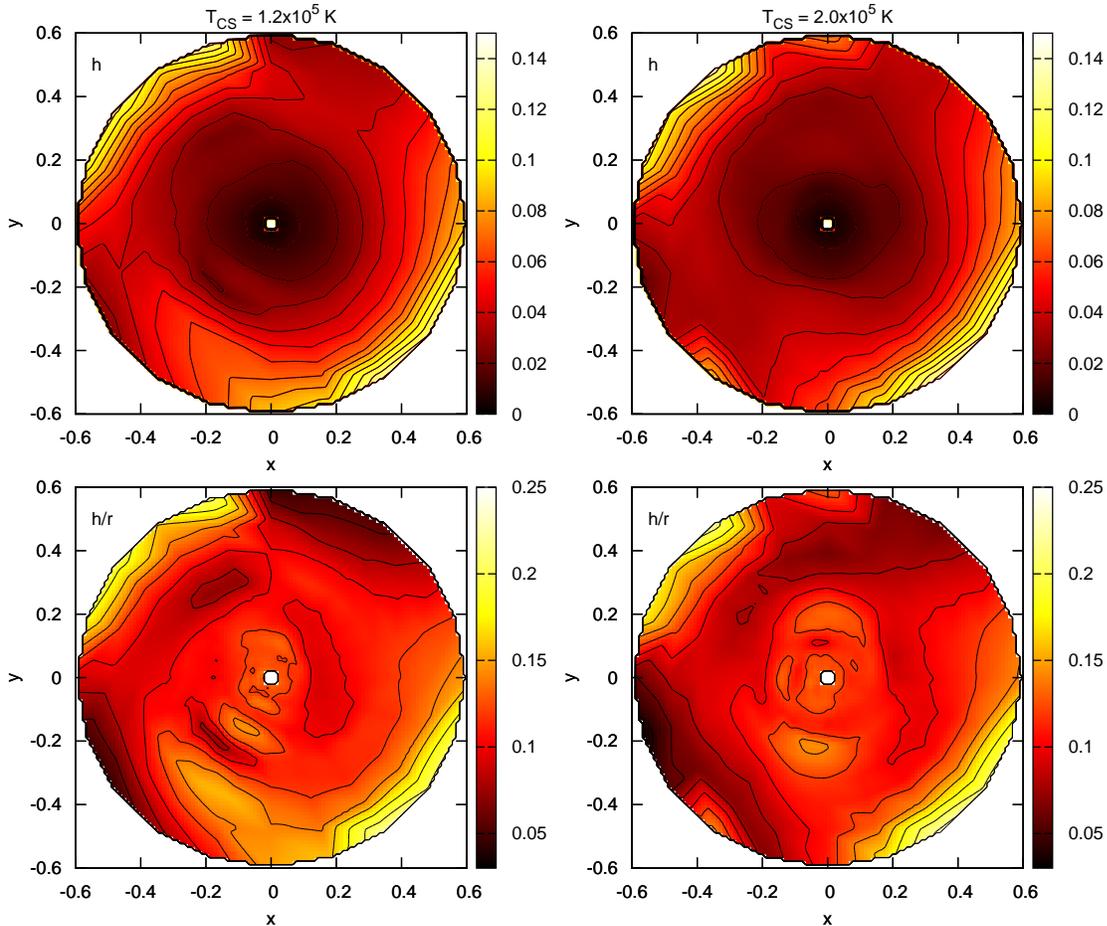}
  \end{center}
  \caption{Height maps of the irradiated disks calculated from the
    data on Day~5 with $T_{\rm cs}=1.2\times 10^5$~K (left) and
    $2.0\times 10^5$~K (right). The upper and lower panels show the
    maps of $h$ and $h/r$, respectively. Both of the contours and
    color-maps represent the same $h$ or $h/r$. The upper panel shows
    contours of $h=0.00$---$0.16$ with an interval of 0.01. The lower
    panel shows contours of $h/h_{\rm disk}=0.00$---$0.26$ with an
    interval of 0.02. The secondary star is located at
    $(x,y)=(1.0,0.0)$.}\label{fig:v455irr_map} 
\end{figure*} 

\begin{figure}
  \begin{center}
    \FigureFile(80mm,80mm){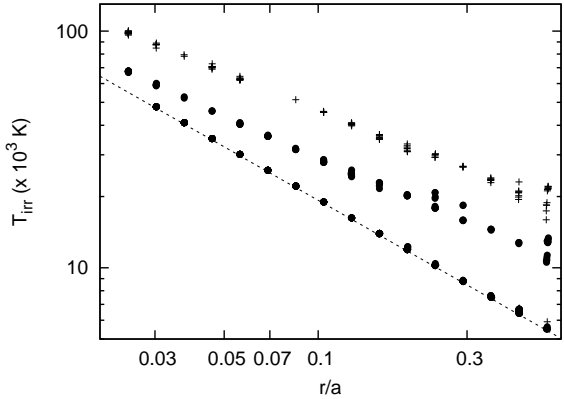}
  \end{center}
  \caption{Radial distribution of temperature of the irradiated
    disk. The filled circles and crosses denote those with $T_{\rm
      cs}=1.2$ and $2.0\times 10^5$~K, respectively. The dashed line
    indicates the standard temperature distribution, that is,
    $T=6.0\times 10^4 (r/r_{\rm min})^{-3/4}$.}\label{fig:temp_irr_r} 
\end{figure}

We calculated the irradiated disk structure using the data on Day~5. 
The resultant height maps are shown in figure~\ref{fig:v455irr_map}.
The left panels are obtained with $\eta=0.5$ and $T_{\rm cs}=1.2\times
10^5$~K, and $T_{\rm in}=6.0\times 10^4$~K. The colors of the
reconstructed disk are consistent with the observation within a color
index of 0.03. The other model parameters are listed in
table~\ref{tab:v455}. Figure~\ref{fig:temp_irr_r} shows the radial 
distribution of temperature of the irradiated disk, as indicated by
the filled circles. We can see two sequences in the temperature; the
high-temperature sequence corresponds to the patches irradiated by the
central source. The low-temperature sequence corresponds to the
non-irradiated patches in which the emission from the central source
is blocked by other patches. A part of the outermost region is heated
up to $1.3\times 10^4$~K by the irradiation. 

The most noteworthy feature of the irradiated disk is that the inner
arm-like patterns are weaker than those estimated from the model
without the irradiation effect. The arm-like pattern has the maximum
$h/r$ of 0.16 in the irradiated disks, although they have $h/r\sim
0.26$ in non-irradiated disks. The weakening of the arm-like pattern
is due to an increase of the temperature of the outermost flaring
regions. As mentioned in subsection~3.2, the inner arm-like pattern in
the left-lower quadrant is responsible for the deep secondary minimum
in the light curve on Day~5. This structure is required to occult the
blue light originated from the inner high-temperature region. However,
in the irradiated disks, the outermost flaring regions significantly
contribute to the blue light. The deep secondary minimum can be
reproduced if that phase corresponds to the position at which the
contribution from those outermost high temperature regions is small. 

On the other hand, the arm-like pattern can be seen even in the
irradiated disk with $T_{\rm cs}=1.2\times 10^5$~K, while it
weakens. This supports the presence of this feature in the disk. The
arm-like pattern almost disappears when we assume a higher $T_{\rm
  cs}$ of $2.0\times 10^5$~K. The right panels in
figure~\ref{fig:v455irr_map} show height maps obtained with this
high $T_{\rm cs}$. However, the temperature of the outermost region
can be quite high, $2.2\times 10^4$~K, and the color of the
reconstructed disk becomes so blue that it is inconsistent with the
observation. 

In addition to the irradiated disk, we evaluate the contribution from
the irradiated secondary star. The contribution is expected to be low
because the irradiated area by the central source is small; 
low-latitude regions are blocked by the disk. We calculated the
irradiation effect of the secondary using the same manner as described
above for the disk. The surface of the secondary was divided into 128
patches (16 and 8 bins in the longitude and latitude directions,
respectively). The temperature was initially set to be 3000~K for all
patches, and calculated with the irradiation effect for each
patch. The model parameters for the disk was the same as the above
case of $T_{\rm cs}=1.2\times 10^5$~K (see table~1). As a result, we
confirmed that the irradiated region is limited in a latitude of
35---80~deg. The maximum temperature is $1.3\times 10^4$~K. The total
flux from the irradiated secondary is $<0.02$ times that from the
irradiated disk. Its contribution is highest at phase~0.5, where 
the irradiated region is facing to us, and lowest at phase~0.0. This
small contribution means that the amplitude of the light curve, which
is caused by the geometrical effect of the disk, is 0.02~mag smaller
than the observed light curve. Compared with the typical photometric
error of 0.01~mag, it leads to a marginal difference in the disk
structure. We subtracted the secondary-star contribution from the
observed light curve, and re-calculated the disk structure. The
resultant height map has a global structure the same as that without the
secondary-star contribution, while the disk height at the outermost
region is slightly smaller. Thus, the irradiated secondary star has no
significant effect on the global structure of the reconstructed height
map. 
 
\subsection{Structure of the Light Curves of Early Superhumps}

\begin{figure}
  \begin{center}
    \FigureFile(80mm,80mm){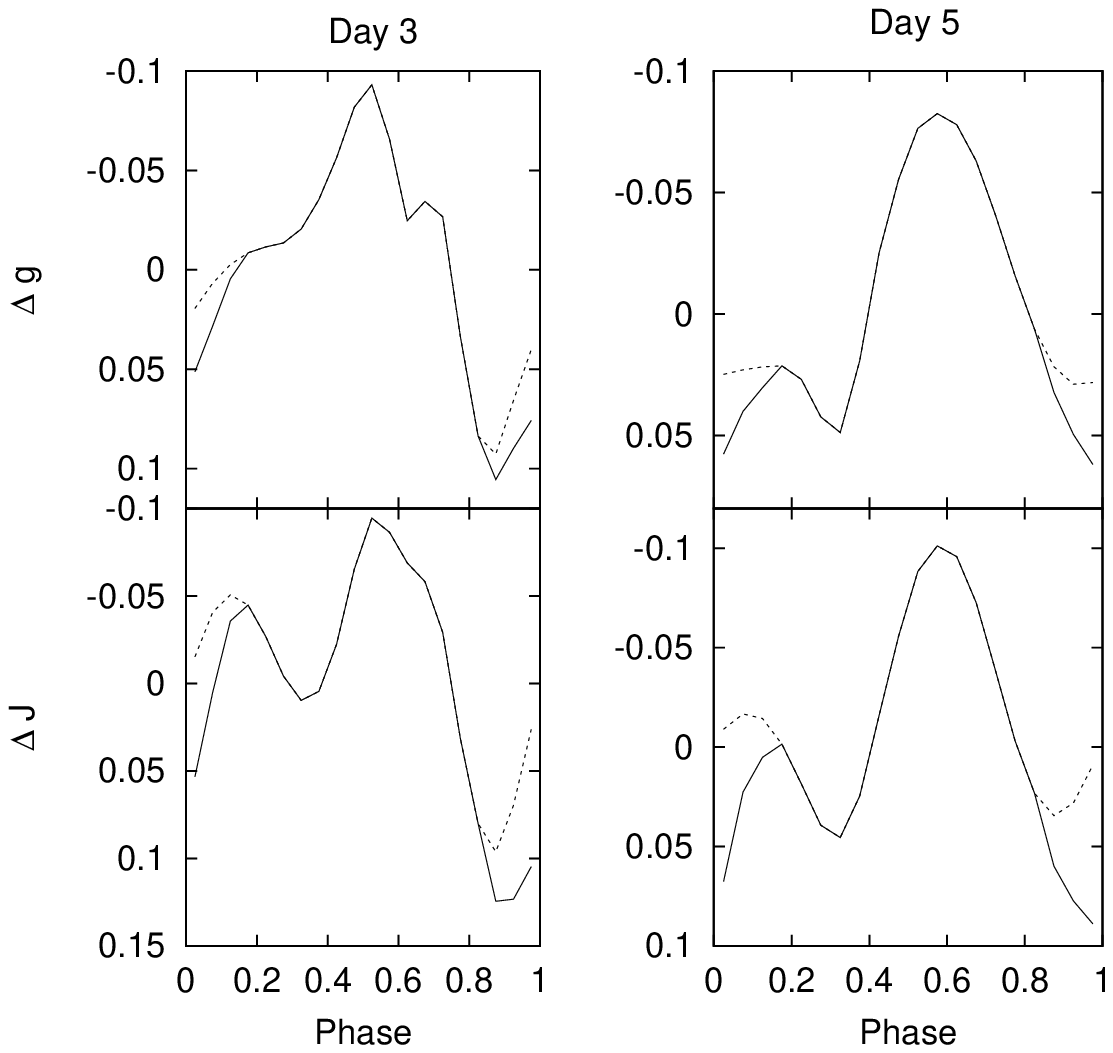}
  \end{center}
  \caption{Model light curves with and without
    eclipses. The left and right panels show those on Day~3 and 5,
    respectively. The upper and lower panels show their light curves
    in the $g$ and $J$-bands, respectively. The eclipsed light
    curves (solid lines) are normalized by the average
    magnitudes in each band. The non-eclipsed light curves (dotted
    lines) are shifted to be consistent with the non-eclipsed light
    curves out of eclipses.}\label{fig:lc_lcecl}
\end{figure} 

Our analysis suggests that the eclipse by the secondary star plays a
crucial role in determining the structure of the light curve around
the primary minimum. Figure~\ref{fig:lc_lcecl} shows 
the $g$- and $J$-band light curves on Day~3 and 5 with (the solid
lines) and without (the dotted lines) eclipses. As can be seen in
these light curves, the eclipses change the phases of the primary
minimum and secondary maximum. It is most prominent in the $J$-band
light curve on Day~5; the phase of the primary minimum is 0.875
if the eclipse is not taken into account, while it is about 0.000 in
the eclipsed light curve. The phases of the secondary maxima are
0.075 and 0.175 in the light curves without and with eclipses,
respectively. The phase shift of the primary minimum indicates that
the phase of the flux minimum can be inconsistent with the 
orbital phase $0$. In fact, the flux minimum preceded the
orbital phase $0$ in all of the light curves on Day~3, as can be seen in
figure~\ref{fig:lc_lcecl}. Hence, the flux minimum cannot
be used for determining the ephemeris of binaries, even in the case of
an eclipsing system.

\begin{figure}
  \begin{center}
    \FigureFile(80mm,80mm){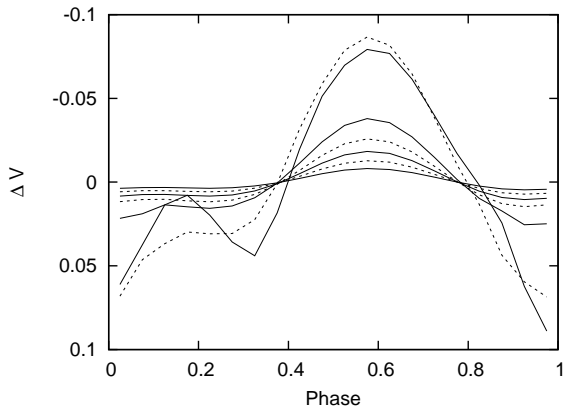}
  \end{center}
  \caption{Model light curves calculated from the height map on Day~5
    for different $i$. From those having the smallest amplitude, the
    solid lines indicate $i=20$, 40, 60, and 80, and the dashed lines
    indicate $i=30$, 50, and 70.}\label{fig:lcecl_incl} 
\end{figure} 

The results of our early superhump mapping can be used to predict the 
amplitude of the light curve depending on the inclination angle,
$i$. Figure~\ref{fig:lcecl_incl} shows the $V$-band model light-curves 
of early superhumps in different $i$, which were calculated by using
the height map on Day~5. We can see a general trend that the light
curves with higher-$i$ have larger
amplitudes. Figure~\ref{fig:lcecl_incl} includes light curves up
to $i=80$~deg. In the case of $i>80$~deg, the amplitude rapidly
increases because of the eclipse of the inner region. It is over 1~mag
when $i>84$~deg. The amplitude of early superhumps is
below 0.10 and 0.01~mag in $i<60$ and $<20$~deg,
respectively. \citet{kat09pdot} report several light curves of early
superhumps, in which a lower limit of the amplitude is apparently
$\sim 0.02$~mag. This can be considered as a detection limit of early
superhumps with typical equipment that they used. The model light
curve has an amplitude smaller than $0.02$~mag in the case of
$i\lesssim 30$~deg. If the direction of the rotation axis of binaries
has a uniform distribution on the sphere, the probability distribution
of systems having $i$ is proportional to $\sin{i}$. Then, the fraction
of systems having $i>30$~deg to all systems is calculated to be
0.87. This implies that early superhumps can be detected in most WZ~Sge
stars. Early superhumps were detected in 22 systems among 34 WZ~Sge
stars which are included in \citet{kat09pdot}. The fraction is 0.65,
which is smaller than the expected one. This is probably because early
stages of superoutbursts are occasionally overlooked. Our result
implies that the fraction of the detection of early superhumps would
increase if early stages are observed more frequently by prompt
reports of outburst discoveries and follow-up time-series observations. 
We note that the above discussion is based on the data on Day~5, on
which the amplitude of the early superhump was smaller than that on
Day~3. The estimated amplitudes and fraction of the detection would be
higher than the above values at the maximum of superoutbursts.

\section{Summary}

We have developed a method by which we can reconstruct the height map of
accretion disks in dwarf novae using multi-band light curves of early
superhumps. Using this ``early superhump mapping'' method, we analyzed
light curves of V455~And during its superoutburst. Our findings are
summarized below.
\begin{itemize}
\item The reconstructed disk has a structure that has two 
  flaring parts in the outermost region and two flaring arm-like
  patterns elongated to relatively inner parts of the disk.
\item The maximum $h/r$ reaches $0.25$ and $0.20$ on Day~3 and
  5, respectively. We suggest that the decrease in the amplitude of
  early superhumps is mainly due to the decrease in the disk height in
  the outermost region. 
\item The overall pattern of the disk is reminiscent of the structure
  of tidally deformed disks. However, one of the inner arm-like
  patterns is difficult to be reproduced by the tidal
  deformation. That pattern is required by the deep secondary minimum
  in the observed light curves. 
\item It suggests that, in addition to the tidal effect, the disk is
  deformed by another unknown mechanism. Alternatively, our model may be
  failing to reproduce the characteristics of the disk in dwarf
  novae. We have confirmed that the disk structure with weaker
  arm-like patterns is optimal in the model including the
  irradiation effect. The strongly irradiated disk, however, gives
  quite blue colors which may conflict the observation.
\item Our result predicts that early superhumps are detected in
  $87$~\% of WZ~Sge stars with amplitudes of $>0.02$~mag. 
\end{itemize}

\bigskip

This work was partly supported by a Grand-in-Aid from the Ministry of
Education, Culture, Sports, Science, and Technology of Japan
(19740104, 22540252). We would like to thank Drs. S. Mineshige, 
D. Nogami, A. Arai, K. Matsumoto, and S. Nakagawa for their comments
and suggestions on this paper. 

\appendix
\section{Details of MCMC algorithm}

We estimated $\{h\}$ by using the MCMC method. Here, we describe details
of the MCMC algorithm and process to obtain $\{h\}$. 

We used the multiple-try Metropolis (MTM) algorithm to sample the
posterior probability distribution of our Bayesian model
(\cite{liu00MTM}). The MTM algorithm is a modified Metropolis-Hastings
algorithm, allowing larger step sizes with a reasonable acceptance
rate. The dimension of our model is equal to the number of the
cross-points of grids, namely $N=N_\theta(N_r+1)$. In the main text, we
set $N_r=16$ and $N_\theta=20$, hence $N=340$. The MTM algorithm has
an advantage in high dimension problems. We used the MTM algorithm
because it gives convergence with a reasonable amount of computer
time, while the standard Metropolis-Hastings algorithm does not. 

We describe the procedure used to make the next set of $h$ ($=\mathbf{y}$)
from a given set of $h$ ($=\mathbf{x}$) in the MTM algorithm in our
model: 1) Draw $k$ trial proposals, $\mathbf{y}_1$, ...,
$\mathbf{y}_k$ from a proposal distribution,
$Q(\mathbf{x},.)$. 2) Select $\mathbf{y}$ among the trial
set $\{ \mathbf{y}_1,..., \mathbf{y}_k \}$ with probability
proportional to the posterior probability density,
$P(\mathbf{y}_i)$ [see, equation~(2)]. 3) Draw $\mathbf{x}_1$, ...,
$\mathbf{x}_{k-1}$ from $Q(\mathbf{y},.)$, and set
$\mathbf{x}_k=\mathbf{x}$. 4) Accept $\mathbf{y}$ with probability, 
\begin{eqnarray}
p=\min \left( 1, \frac{P(\mathbf{y}_1)+ ... +P(\mathbf{y}_k)}{P(\mathbf{x}_1)+ ... +P(\mathbf{x}_k)}  \right),
\end{eqnarray}
and reject it with probability $1-p$. In the present study, we set $k=5$.
The proposal distribution $Q(\mathbf{y},\mathbf{x})$ is a multivariate
normal distribution with a mean of $\mathbf{x}$ and a covariance
matrix of $s\Sigma$. We defined $\Sigma$ as the covariance matrix of
the probability distribution, $\pi'$, which is an approximated
probability of the prior, $\pi$; $\pi'=\pi_{\rm smooth}
\pi'_{\rm disk}$, where $\pi'_{\rm disk}$ is $\pi_{\rm disk}$ in $h>0$ [see,
equations~(4) and (5)]. We control the step size by the scalar parameter,
$s$. We update $h$ of one quadrant of the disk in one step, and all $h$ is
updated in four steps. We choose $s$ to achieve the acceptance rate of
one step close to $0.1$. 

\begin{figure}
  \begin{center}
    \FigureFile(80mm,80mm){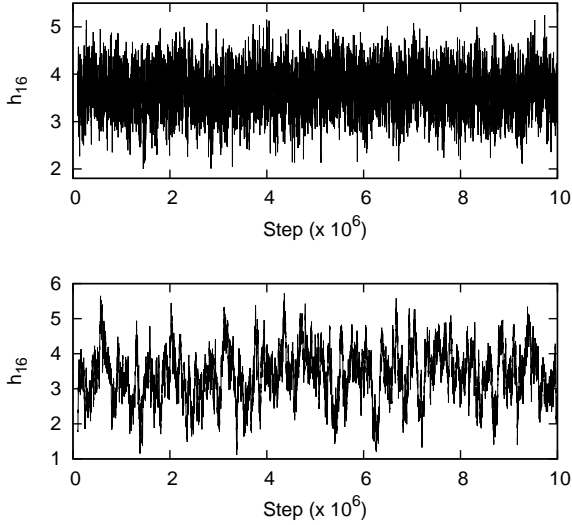}
  \end{center}
  \caption{Traces of a height parameter in the MCMC run. The
    calculations were performed with the same parameter as the case of
  Day~5 (see, table~1) with $w=1.0$ in the upper panel and $2.0$ in
  the lower panel.}\label{fig:conv} 
\end{figure} 

\begin{figure}
  \begin{center}
    \FigureFile(80mm,80mm){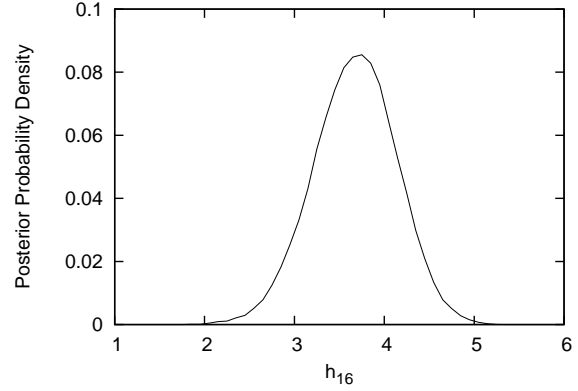}
  \end{center}
  \caption{Posterior probability density function of one of height
    parameters, $h_{16}$.}\label{fig:post_v455d5}  
\end{figure} 

In all MCMC calculation presented in this paper, we discard the first
$10^5$ steps as a MCMC burn-in phase. All chains reached convergence
after the burn-in phase. An example can be seen in 
the upper panel of figure~\ref{fig:conv}, which shows the traces of a
height parameter, $h_{16}=h(0.49,0.00)$ in the MCMC run in
subsection~3.2. We performed $10^7$ steps and made the height map. We
confirmed that there was no significant difference among the
reconstructed height maps made from sub-samples of $10^6$ steps.
Figure~\ref{fig:post_v455d5} shows the posterior probability density
function of $h_{16}$. The distribution has a single peak, and there is
no evidence of other peaks. The form of the distribution indicates
that the solution of the height map is uniquely determined.

\begin{figure}
  \begin{center}
    \FigureFile(80mm,80mm){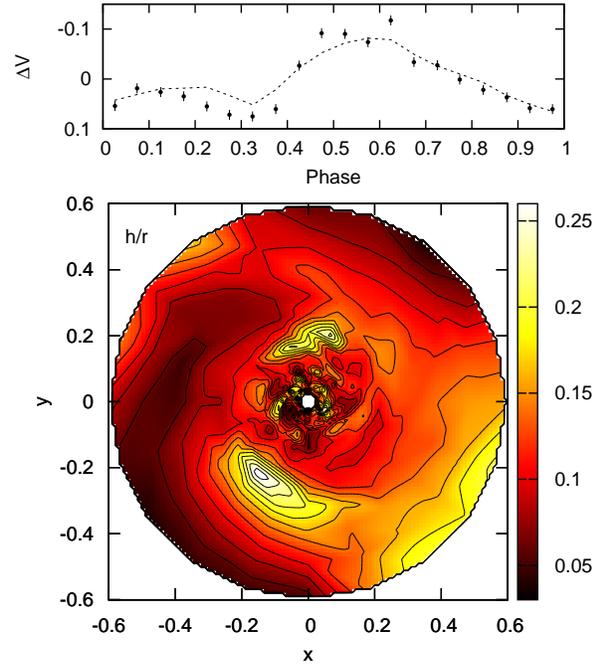}
  \end{center}
  \caption{Light curve (upper) and height map (lower) calculated with
    $w=2.0$ using the data on Day~5 of V455~And. In the upper
    panel, the filled circles represent the observed light curves. The
    dashed line indicates the model light curve. The color maps and
    contours represent $h/r$. The contours are shown from 
    $h/r=0.00$---$0.26$ with an interval of 0.02. The secondary star is
    located at $(x,y)=(1.0,0.0)$.}\label{fig:v455d5w20}  
\end{figure} 

All calculations in the present paper were performed with $w=1.0$. The
trace of $h_{16}$ calculated with $w=2.0$ is shown in the lower panel
of figure~\ref{fig:conv}. In this case, it takes $>10$ times longer
than the case of $w=1.0$ until convergence. Furthermore, the MCMC
mixing is slow, as can be seen in figure~\ref{fig:conv}. As a result,
in order to obtain the final result, the amount of calculations with
$w=2.0$ is much larger than that with $w=1.0$, and was impractical for
the present study. It is an issue of computational time. It is
noteworthy that, even with $w=2.0$, 
the global pattern of the reconstructed height maps is consistent with
that reported in section~3. Figure~\ref{fig:v455d5w20} shows the
results calculated with $w=2.0$: the model light curve (upper panel)
and the reconstructed $h/r$ map (lower panel). They were obtained from
the last $5\times 10^6$ steps in the lower panel of
figure~\ref{fig:conv}. The outermost and inner arm-like flaring
patterns can be seen in the $h/r$ map in figure~\ref{fig:v455d5w20},
as seen in the case with $w=1.0$ (figure~\ref{fig:v455d5map}). The
flaring regions in figure~\ref{fig:v455d5w20} tend to be more
localized than those in figure~\ref{fig:v455d5map}, as expected from
the higher $w$. It indicates that a high $w$ plays a role in
increasing the resolution of the reconstructed height map, while the
global structure of the map remains. In the main text, we used $w=1.0$
because we have focused not on sub-structures, but on the global
deforming-pattern of the disk.


\end{document}